\documentclass[pra,twocolumn,a4paper]{revtex4}
\usepackage{bm,graphicx,amsmath}

\usepackage{graphicx}
\usepackage{hyperref}
\usepackage{epstopdf}
\graphicspath{{figures/}}

\begin{document}

\title{Creating Single Collective Atomic Excitations via Spontaneous Raman Emission in
Inhomogeneously Broadened Systems : Beyond the Adiabatic
Approximation}
\author{Carlo Ottaviani$^{1,2}$, Christoph Simon$^1$, Hugues de Riedmatten$^1$, Mikael Afzelius$^1$,
Bj\"{o}rn Lauritzen$^1$, Nicolas Sangouard$^{1,3}$ and
Nicolas Gisin$^1$} \affiliation{$^1$ Group of Applied
Physics, University of Geneva, Switzerland.\\$^2$
Departament de F\'isica, Universitat Aut\'onoma de
Barcelona, E-08193 Bellaterra, Spain.\\$^3$ Laboratoire
Mat\'{e}riaux et Ph\'{e}nom\`{e}nes Quantiques CNRS,
UMR7162, Universit\'{e} Paris Diderot, France.}

\date{\today}

\begin{abstract}
The creation of single collective excitations in atomic
ensembles via spontaneous Raman emission plays a key role
in several quantum communication protocols, starting with
the seminal DLCZ protocol [L.-M.Duan, M.D. Lukin, J.I.
Cirac, and P. Zoller, Nature {\bf 414}, 413 (2001).] This
process is usually analyzed theoretically under the
assumptions that the write laser pulse inducing the Raman
transition is far off-resonance, and that the atomic
ensemble is only homogeneously broadened. Here we study the
impact of near-resonance excitation for inhomogeneously
broadened ensembles on the collective character of the
created atomic excitation. Our results are particularly
relevant for experiments with hot atomic gases and for
potential future solid-state implementations.
\end{abstract}

\maketitle

\section{Introduction}

\begin{figure}
    \center
  \includegraphics[angle=0
  ,
  width=0.6 \columnwidth]{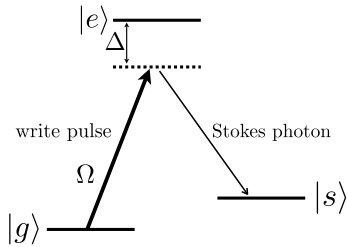}\\
  \caption{Basic level scheme for the creation of collective atomic
  excitations in atomic ensembles via spontaneous Raman emission. All atoms start
  out in $g$. A laser
  pulse off-resonantly excites the $g-e$ transition, making it possible for
  a photon to be emitted on the $e-s$ transition (with small probability).}
\label{levels}
\end{figure}

The creation of collective atomic excitations via
spontaneous Raman emission plays an important role in
well-known quantum information protocols, in particular the
quantum repeater \cite{briegel} protocol proposed by Duan,
Lukin, Cirac and Zoller (DLCZ) \cite{DLCZ}; for other
protocols based on the same process see
\cite{jiang,zhao,chen,IPP}. The basic (idealized) scheme is
as follows, cf. Fig. \ref{levels}. In an ensemble of
three-level systems with two ground states $g$ and $s$ and
an excited state $e$ all $N$ atoms are initially in the
state $g$. An off-resonant laser pulse on the $g-e$
transition (the {\it write} pulse) leads to the spontaneous
emission of a Raman photon on the $e-s$ transition (the
{\it Stokes} photon). Detection of this photon in the far
field, such that no information is revealed about which
atom it came from, creates an atomic state that is a {\it
coherent} superposition of all the possible terms with
$N-1$ atoms in $g$ and one atom in $s$, in the simplest
case the completely symmetric state
\begin{equation}
\frac{1}{\sqrt{N}}(|s\rangle|g\rangle...|g\rangle+|g\rangle|s\rangle|g\rangle
...|g\rangle+...+|g\rangle...|g\rangle|s\rangle).
\label{ideal}
\end{equation}
Such a state corresponds to a single {\it collective atomic
excitation} in $s$. In general the term with the $n$-th
atom in $s$ will have a phase $e^{i({\bf k}_w-{\bf k}_S)
{\bf x}_n}$, where ${\bf k}_w$ is the ${\bf k}$ vector of
the write laser, ${\bf k}_S$ is the ${\bf k}$ vector of the
detected Stokes photon, and ${\bf x}_n$ is the position of
the $n$-th atom. The phases in Eq. (\ref{ideal}) thus
correspond to the case ${\bf k}_S={\bf k}_w$. Moreover in
practice the amplitudes of the different terms may vary,
depending on the laser profile and the shape of the atomic
ensemble.

A remarkable feature of such collective excitations that is
of great interest for practical applications is that they
can be {\it read out} very efficiently by converting them
into single photons that propagate in a well-defined
direction, thanks to collective interference
\cite{DLCZ,laurat,simon-vuletic}. Resonant laser excitation
of such a state on the $s-e$ transition (the {\it read}
laser pulse) leads to an analogous state with $N-1$ atoms
in $g$ and one delocalized excitation in $e$. All the terms
in this state can decay to the initial state
$|g\rangle^{\otimes N}$ while emitting a photon on the
$e-g$ transition (the {\it Anti-Stokes} photon). If the
phase matching condition ${\bf k}_S+{\bf k}_{AS}={\bf
k}_w+{\bf k}_r$ is fulfilled, where ${\bf k}_r$ is the
${\bf k}$ vector of the read laser and ${\bf k}_{AS}$ that
of the Anti-Stokes photon, then the amplitudes
corresponding to the various terms interfere constructively
(provided that there are no other effects disturbing the
interference, such as atomic motion or the effects studied
in this paper, cf. below), leading to a very large
probability amplitude for emission of the Anti-Stokes
photon in the direction given by ${\bf k}_w+{\bf k}_r-{\bf
k}_S$. For atomic ensembles that contain sufficiently many
atoms, emission in this one direction can completely
dominate all other directions. This allows a very efficient
collection of the Anti-Stokes photon
\cite{laurat,simon-vuletic}.

Note that there is no such collective interference effect
for the emission of the Stokes photon, since its emission
by different atoms corresponds to orthogonal final states,
e.g. the state $|s\rangle |g\rangle...|g\rangle$ if the
Stokes photon was emitted by the first atom etc. Full
``which-way'' information about the origin of the photon is
thus stored in the atomic ensemble, making interference
impossible \cite{scullyscience}. As a consequence the total
emission probability for the Stokes photon is simply given
by the sum of the emission probabilities for each atom, and
there is no preferred direction of emission.

The creation of collective excitations via spontaneous
Raman emission is usually analyzed theoretically under the
assumption that the write pulse is far off-resonance
\cite{DLCZ,DCZ} (but see Section VI.B of Ref.
\cite{kolchinpra}) . Under this condition it is possible to
adiabatically eliminate the excited state. However, in
experiments the far off-resonance condition is frequently
not fulfilled. From an experimental point of view it can be
advantageous to approach resonance in order to increase the
rate for the spontaneous Raman process or in order to avoid
exciting nearby levels. For excitation relatively close to
resonance it is no longer justified to eliminate the
excited state. The precise frequency of the excited state
then influences the dynamics \cite{courtens}, and it
becomes important to consider the effects of inhomogeneous
broadening of the transition between ground and excited
state \cite{homogeneous}, which is significant in many
experimental situations, e.g. for hot gases
\cite{VanderWal,EisamanPRL,EisamanNature,Manz}, where the
relevant mechanism is Doppler broadening. (Note that the
effects of Doppler broadening are negligible in similar
experiments with cold atomic gases
\cite{laurat,simon-vuletic,coldgases}.)

Inhomogeneous broadening will also be an essential factor
in future experiments with solid-state atomic ensembles, in
particular rare-earth doped crystals \cite{EisamanRE},
where it is due to the crystal environment. These systems
are otherwise very attractive candidates for realizing the
DLCZ and similar protocols thanks to their excellent
coherence properties. For example, storage times exceeding
1 second have already been demonstrated in such a system
for coherent atomic excitations in $s$ created via
electromagnetically induced transparency \cite{RE-EIT}, and
light at the single-photon level has been stored and
re-emitted using the ``atomic frequency comb'' protocol
\cite{DeRiedmatten,Afzelius}. Ref. \cite{EisamanRE}
proposed to reduce the inhomogeneity in these solid-state
systems via spectral tailoring techniques similar to those
employed in light storage experiments
\cite{RE-EIT,DeRiedmatten,sellarsCRIB}. However, such an
approach greatly reduces the number of available atoms,
making it much harder to write and read the atomic
excitations efficiently.

Motivated by these considerations, we here analyze the
creation of collective atomic excitations in inhomogeneous
systems by spontaneous Raman emission without resorting to
the usual adiabatic elimination of the excited state. This
makes it possible to quantify the impact of near-resonance
excitation in combination with inhomogeneous broadening on
the {\it collectivity} of the created atomic excitation. We
introduce the term collectivity for the fidelity of the
created excitation with respect to the ideal state of Eq.
(\ref{ideal}). This quantifies the degree of collective
interference that is possible when reading out a given
atomic excitation. It is equal to one (corresponding to the
possibility of perfect collective interference) for
excitations that are created under far off-resonant
conditions. It is reduced for near-resonant excitation in
inhomogeneous systems. In such systems, atoms closer to
resonance with the write laser will have a larger amplitude
of emitting a Raman photon. Moreover for near-resonant
(non-adiabatic) excitation the excited state plays a role
in the Raman process, leading to phases that differ from
atom to atom in inhomogeneous systems. These effects
perturb the collective interference that is at the heart of
the read-out process.

The main goal of our present work is the characterization
of the collective atomic excitation that is created by the
emission and detection of the Stokes photon, with a focus
on {\it spectral} aspects due to near-resonant write
excitation and inhomogeneous broadening. Spatial effects
have previously been analyzed in detail in Ref. \cite{DCZ},
and propagation effects for the Anti-Stokes photon in Ref.
\cite{kolchinpra}. This paper is organized as follows. In
section II we study spontaneous Raman emission for a single
atom under pulsed excitation. We show that the amplitude
for detecting a Stokes photon at a given time is
proportional to the amplitude of the atom being in the
excited state. In section III we show how the single-atom
results can be used to quantify the collectivity for an
inhomogeneously broadened atomic ensemble. In section IV we
give numerical examples relevant to hot atomic gases and
solid-state systems. Section V contains our conclusions and
an outlook towards future work.

\section{Spontaneous Raman emission for a single atom under pulsed excitation}

\begin{figure}
    \center
  \includegraphics[angle=0
  ,
  width=0.8 \columnwidth]{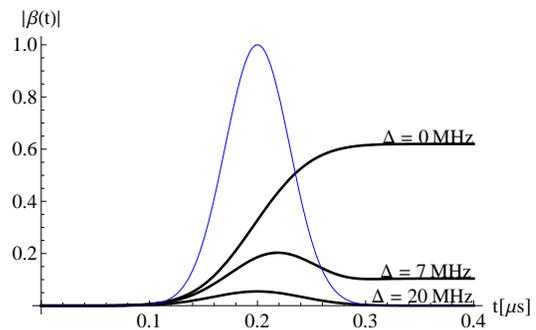}\\
  \caption{Evolution of the absolute value of the excited state amplitude
  for a single atom (that starts out in the ground state) under pulsed excitation for different values of detuning
  $\Delta$. The fine line shows the exciting laser pulse $\Omega(t)$,
  which is a Gaussian centered at 0.2 $\mu$s with FWHM pulse duration
  of 0.1 $\mu$s and a maximum Rabi frequency of 1 MHz. The same pulse is
  used in the examples throughout the paper.} \label{single}
\end{figure}

Let us first recall the well-known dynamics of a two-level
system with levels $e$ and $g$ under pulsed excitation, cf.
Fig. \ref{single}. Suppose that the system starts out in
$g$ and that it is excited by a pulse with Rabi frequency
$\Omega(t)$ that is detuned from resonance by $\Delta$. We
focus on the case $\int dt \Omega(t)<1$ (moderate pulse
area). Depending on the value of $\Delta$ there are
different regimes. For far off-resonance excitation,
$\Delta \gg \Omega(t)$, the amplitude to be in $e$ is well
approximated by $-\frac{\Omega(t)}{\Delta}$, such that the
system returns to $g$ after the pulse. This is called the
adiabatic regime, where the system always stays in the
momentary lowest-energy eigenstate of the Hamiltonian, cf.
the lowest curve in Fig. \ref{single}. For smaller values
of $\Delta$ the population in $e$ attains a maximum during
the pulse and then declines, but without returning exactly
to zero, cf. the intermediate curve in Fig. \ref{single}.
Finally, for resonant excitation the population in $e$
increases monotonously for the whole duration (as long as
one can neglect the spontaneous decay of the excited
level), cf. the top curve in Fig. \ref{single}.

Now add a third level $s$. We are interested in spontaneous
emission on the $e-s$ transition. This means that we have
to include a continuum of vacuum modes of the
electro-magnetic field. We follow the Wigner-Weisskopf
treatment of spontaneous emission \cite{scullybook}. The
relevant states are now $|g\rangle|0\rangle$,
$|e\rangle|0\rangle$, where $|0\rangle$ is the vacuum state
of the field, and the set $|s\rangle |1_{\omega}\rangle$,
where $|1_{\omega}\rangle$ are a continuum of one-photon
states labeled by the frequency $\omega$. For simplicity,
and motivated by typical experiments, we focus on one
specific direction of emission, treating the effect of the
other modes globally as a contribution to the decay of the
excited level $e$; note that $e$ can moreover also decay on
the $e-g$ transition.

Denoting the amplitudes of the states $|g\rangle|0\rangle$,
$|e\rangle|0\rangle$ and $|s\rangle |1_{\omega}\rangle$ by
$\alpha$, $\beta$ and $\gamma_{\omega}$ respectively, and
the overall decay rate of $e$ by $\Gamma$, the dynamical
equations are
\begin{eqnarray}
\dot{\alpha}=-i\Omega \beta \\
\dot{\beta}=-i\Omega \alpha - i\Delta \beta-\Gamma \beta-ig
\sum_{\omega}
\gamma_{\omega}\\
\dot{\gamma}_{\omega}=-ig\beta-i\omega\gamma_{\omega}.
\label{dynamics}
\end{eqnarray}
The coupling constant $g$ depends on the dipole moment of
the $e-s$ transition, but also on the quantization volume
and on the solid angle of the mode under consideration. Its
precise form is not important for our purposes here, cf.
Ref. \cite{scullybook}. The solution of Eq.
(\ref{dynamics}) is
\begin{equation}
\gamma_{\omega}(t)=-ig\int_0^t dt'
e^{-i\omega(t-t')}\beta(t'). \label{gammaomega}
\end{equation}
Inserting this into Eq. (3) gives
\begin{equation}
\dot{\beta}=-i\Omega \alpha - i\Delta \beta-\Gamma
\beta-g^2 \sum_{\omega}\int_0^t dt'
e^{-i\omega(t-t')}\beta(t').
\end{equation}
Changing the order of summation and integration in the last
term, and making a Wigner-Weisskopf type approximation,
\begin{equation}
\sum_{\omega} e^{-i\omega(t-t')} \propto \delta(t-t'),
\label{ww}
\end{equation}
we see that the modes under consideration just make a
contribution to the overall decay term $-\Gamma \beta$, as
was to be expected. The relative size of this contribution
depends on the solid angle of the considered mode (and also
on the branching ratio between the $e-g$ and $e-s$
transitions). It is essentially negligible in typical
experimental situations where the solid angle of collection
is small. More interesting for our purposes is the effect
of the detection of a Stokes photon at time $t$. Since the
annihilation operator for the mode under consideration
satisfies $a=\sum_{\omega} a_{\omega}$ (in the
Schr\"{o}dinger picture), the amplitude for such a
detection is
\begin{equation}
c(t)=\sum_{\omega}
\gamma_{\omega}(t)=-\sum_{\omega}ig\int_0^t dt'
e^{-i\omega(t-t')}\beta(t')\propto \beta(t),
\label{cpropbeta}
\end{equation}
where the equality follows from Eq. (\ref{gammaomega}) and
the proportionality from Eq. (\ref{ww}). This shows that
the amplitude for detecting a Stokes photon is simply
proportional to the amplitude of the atom being in the
excited state. Below we will show that, as a consequence,
the (numerical) solution of the two-level problem gives us
all the information we need in order to study the
collective interference. For a single atom, the atomic
state conditional on detecting a Stokes photon is simply
$|s\rangle$ in the single-atom case. It is more interesting
in the $N$-atom case below.

\section{Collectivity in inhomogeneously broadened systems}

We now consider the situation where the laser pulse excites
an ensemble of $N$ atoms, which do not all have the same
resonance frequency. Without the third level $s$ one would
have a time-dependent state
\begin{equation}
\prod \limits_{n=1}^N
(\alpha_n(t)|g\rangle_n+\beta_n(t)|e\rangle_n)\equiv \prod
\limits_{n=1}^N |G_n(t)\rangle.
\end{equation}
We are interested in the case where there is a third level,
but where the spontaneous emission of a photon on the $e-s$
transition into the considered directional mode occurs with
only a small probability. We then take into account only
terms that correspond to a single emission. Detection of a
single Stokes photon at time $t$ creates a conditional
state proportional to
\begin{equation}
c_1(t)|s\rangle|G_2(t)\rangle...|G_N(t)\rangle+...+c_N(t)|G_1(t)\rangle...|G_{N-1}(t)\rangle|s\rangle,
\label{Nstate}
\end{equation}
where for simplicity we again assume ${\bf k}_S={\bf k}_w$
(no phase factors). The key point is that the coefficients
$c_n(t)$ in Eq. (\ref{Nstate}) are given by the single-atom
calculation described in the previous section. Eq.
(\ref{cpropbeta}) shows that $c_n(t)$ is in fact
proportional to $\beta_n(t)$, which depends on the detuning
of the corresponding ($n$-th) atom with respect to the
laser. The collective atomic state is thus proportional to
\begin{equation}
\beta_1(t)|s\rangle|G_2(t)\rangle...|G_N(t)\rangle+...+\beta_N(t)|G_1(t)\rangle...|G_{N-1}(t)\rangle|s\rangle.
\end{equation}

After the detection of the considered Stokes photon the
states $|G_n(t)\rangle$ will continue to evolve, and
further photons will be emitted into other directional
modes (let us recall that we are interested in the regime
where the probability to emit another Stokes photon into
the same mode is small). However, in the readout process
the corresponding atomic excitations in $s$ will just lead
to the emission of additional Anti-Stokes photons in other,
undetected directions. As long as the total number of
excitations created in $s$ is much smaller than the total
number of atoms $N$, this has no significant effect on the
collective interference in the readout process.

Neglecting the additional excitations in $s$ discussed in
the previous paragraph, the starting state for the readout
is given by
\begin{eqnarray}
\beta_1(t_S)|s\rangle|G_2(t_M)\rangle...|G_N(t_M)\rangle+...\nonumber\\
+\beta_N(t_S)|G_1(t_M)\rangle...|G_{N-1}(t_M)\rangle|s\rangle,
\end{eqnarray}
where $t_S$ is the time when the Stokes photon was
detected, whereas $t_M$ is the memory time, i.e. the time
when the excitation is read out, which may be much larger
than $t_S$. For long enough $t_M$, depending on the
lifetime of $e$, all the states $|G_k\rangle$ will
essentially be equal to $|g\rangle$. The state of the
atomic excitation in $s$ is then proportional to
\begin{equation}
\beta_1(t_S)|s\rangle|g\rangle...|g\rangle+...
+\beta_N(t_S)|g\rangle...|g\rangle|s\rangle. \label{real}
\end{equation}
We define the {\it collectivity} $C$, which is a function
of the time of emission of the Stokes photon $t_S$, as the
fidelity of the (normalized) state of Eq. (\ref{real}) with
respect to the ideal state of Eq. (\ref{ideal}).

\section{Numerical Results and Discussion}

\begin{figure}
    \center
  \includegraphics[angle=0
  ,
  width=0.7 \columnwidth]{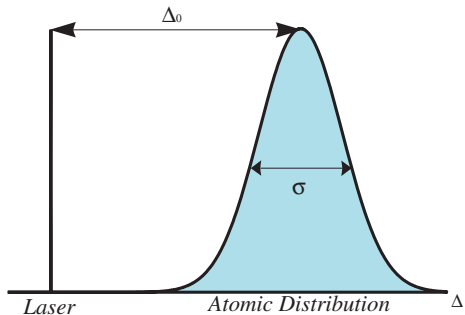}\\
  \caption{We consider the situation where the energy of
  the excited state $e$ is not the same for all the atoms
  in the ensemble, but has a Gaussian distribution with FWHM $\sigma$.
  We have chosen $\sigma=500$ MHz in all our examples.
  We denote the detuning of the write laser with respect to the center of
  this distribution by $\Delta_0$, which varies between 0 and 1.25 GHz in our
  examples.} \label{Delta0sigma}
\end{figure}

We will now consider concrete examples. It is convenient to
approximate the distribution of frequencies of the $N$
atoms by a continuous spectral distribution $n(\Delta)$
with $\int d\Delta n(\Delta)=N$. We will assume a Gaussian
distribution with FWHM $\sigma$ for simplicity. We denote
by $\Delta_0$ the detuning of the laser with respect to the
center of the atomic distribution, see Fig.
\ref{Delta0sigma}.

The collectivity is then defined as
\begin{equation}
C(t_S)=\frac{|\int d\Delta n(\Delta)
\beta(t_S,\Delta)|^2}{N \int d\Delta n(\Delta)
|\beta(t_S,\Delta)|^2},
\end{equation}
where $t_S$ is the time of emission of the Stokes photon as
before. The relevant times are therefore those where Stokes
photon emission is likely. As we have seen before, the
Stokes emission probability for an atom with detuning
$\Delta$ at time $t$ is proportional to the population in
the excited state, $|\beta(t,\Delta)|^2$. Let us recall
that the total emission probability for the Stokes photon
can be obtained by summing this quantity over all atoms
(there is no collective interference for the Stokes photon
emission). It is thus of interest to consider the average
excited state population
\begin{equation}
p_e(t)=\frac{1}{N}\int d\Delta n(\Delta)
|\beta(t,\Delta)|^2.
\end{equation}
The values of $t_S$ that are likely to be observed are
those for which $p_e(t_S)$ is significant.

For the inhomogeneous broadening we choose $\sigma=0.5$
GHz, which is a realistic value both for hot gases
\cite{VanderWal,EisamanPRL,EisamanNature,Manz,Kash} and for
rare-earth doped crystals \cite{Macfarlane}. We furthermore
choose $\Omega(t)$ to have Gaussian temporal shape with a
FWHM pulse duration of 0.1 $\mu$s, centered at $t=0.2 \mu$s
in the figures, and with a maximum Rabi frequency of 1 MHz
(corresponding to $2 \pi \times 10^6 \mbox{rad}/s$), a
choice that is again motivated both by existing hot gas
experiments and by potential future experiments on
rare-earth doped solids.

\begin{figure}
    \center
  \includegraphics[angle=0,  width=0.9 \columnwidth]{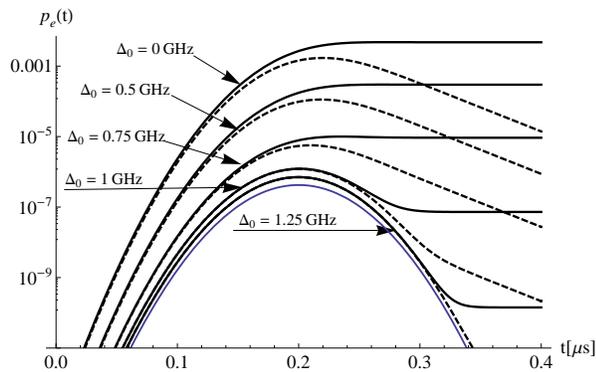}\\
  \caption{The average excited state population $p_e$ as a function of time for different values of
  $\Delta_0$. The probability of emitting a Stokes
  photon is proportional to this quantity. The detuning
  $\Delta_0$ of the laser from the center of the atomic
  distribution
  takes the values 0, 0.5 GHz, 0.75 GHz, 1.0 GHz and 1.25 GHz from top to bottom.  The solid lines correspond to a situation without
spontaneous decay,
  the dashed lines to a decay rate $\Gamma=5$ MHz. The write pulse is the same as
  in Fig. \ref{single}. The blue bottom-most line shows the profile of the square of the Rabi frequency, $\Omega^2(t)$.
  One can see that $p_e$ follows $\Omega^2$ for large $\Delta_0$, in analogy
  with the single-atom situation of Fig. \ref{single}.}
 \label{pe}
\end{figure}

\begin{figure}
    \center
  \includegraphics[angle=0
  ,
  width= 0.9 \columnwidth]{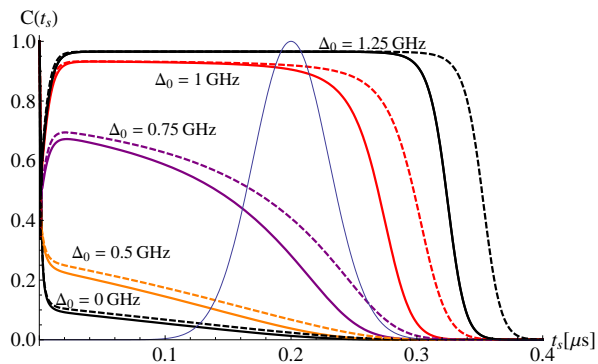}\\
  \caption{Collectivity of the created atomic excitation in
  $s$ as a function of the time of emission of the Stokes
  photon $t_S$ for different values of $\Delta_0$,
  for $\Gamma=0$ (full lines) and $\Gamma=5$ MHz (dashed
  lines). The pulse and inhomogeneous distribution are the
  same as in Fig. \ref{pe}. The square of the Rabi frequency $\Omega^2(t)$ is shown as
  a fine blue line. One can see that a detuning $\Delta_0=1.25$ GHz is
  sufficient to have excellent collectivity for all relevant values
  of $t_S$. } \label{C}
\end{figure}

\begin{figure}
    \center
  \includegraphics[angle=0
  ,
  width= 0.9 \columnwidth]{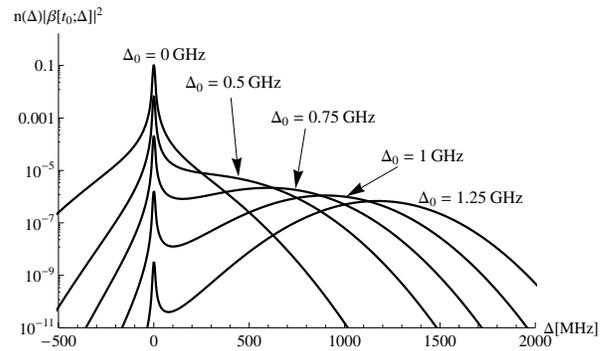}\\
  \caption{The quantity $n(\Delta)|\beta(t_0,\Delta)|^2$ at $t_0=0.2 \mu$s
  (at the center of the pulse) as a function
  of $\Delta$, showing which atomic frequencies contribute significantly to the
  emission of the Stokes photon,
  and thus to the atomic excitation created in $s$. The laser frequency
  is defined to be $\Delta=0$. The center frequency of the atomic distribution
  takes the same values as in Figs. \ref{pe} and \ref{C}. The contribution from the resonant
  atoms (fluorescence)
  dominates for small values of $\Delta_0$, but becomes negligible compared
  to the contribution from the bulk of the atomic distribution (Raman scattering)
  for large $\Delta_0$.} \label{nbeta2}
\end{figure}

Figs. \ref{pe} and \ref{C} show $p_e(t)$ and $C(t_S)$
respectively, for different values of $\Delta_0$. One sees
the transition from the far-detuned (adiabatic) regime,
where the population follows the Rabi pulse and the
collectivity is high for emission times during the duration
of the pulse, to the resonant regime, where the population
no longer follows the pulse and the collectivity is low.
The solid lines in the figures correspond to the case
$\Gamma=0$ (negligible atomic decay), which is realistic
for rare-earth doped solids, where typical excited state
lifetimes are of order 100 $\mu$s to 10 ms
\cite{Macfarlane}. The dashed lines correspond to
$\Gamma=5$ MHz, which is a typical value for hot gases. One
sees that the collectivity is very similar in both cases,
but it stays high a little longer after the pulse in the
case with spontaneous decay. Intuitively, the decay of the
collectivity after the pulse is due to the fact that Stokes
emission for these times is dominated by non-adiabatic
contributions (fluorescence) \cite{courtens}. The
spontaneous decay suppresses these contributions compared
to the adiabatic ones (Raman scattering) and thus enhances
the collectivity. For very early times ($t_S < 0.02 \, \mu
s$) there are oscillations in the collectivity which we
ascribe to a beating between fluorescence and Raman
scattering. However, the Stokes photon emission probability
for these times is exceedingly low, cf. Fig. \ref{pe}.

In the absence of other effects the collectivity of a given
atomic excitation in $s$ would remain unchanged once the
excitation has been created via the emission of the Stokes
photon. In practice it will decay on a timescale given by
the spin coherence time characterizing the atomic ensemble,
which can be very long for solid-state ensembles
\cite{RE-EIT}. The spin transition $g-s$ can also be
inhomogeneously broadened in these systems, however the
associated dephasing can be compensated using spin echo
techniques, which have already allowed the demonstration of
coherence times as long as 30 seconds
\cite{sellarscoherence}. In the case of hot gases the
storage time is also limited by the motion of the atoms
\cite{VanderWal,EisamanPRL,EisamanNature}.

Fig. \ref{nbeta2} shows the quantity $n(\Delta)
|\beta(t_0,\Delta)|^2$ as a function of $\Delta$ for
$t_0=0.2 \mu$s, i.e. at the center of the pulse. This
permits one to see which frequency classes of atoms
contribute significantly to the Stokes emission, and thus
to the collective atomic excitation in $s$ that is created,
for the different values of $\Delta_0$. One sees that for
small $\Delta_0$ only the atoms that are resonant with the
laser contribute significantly. On the other hand for large
$\Delta_0$ the broad contribution from the bulk of the
atomic distribution becomes much more important than that
of the resonant atoms (because there are so few of the
latter). This is consistent with the results for the
collectivity shown in Fig. \ref{C}. Intuitively, the sharp
feature on resonance corresponds to fluorescence, whereas
the broad off-resonant contribution corresponds to Raman
scattering \cite{courtens}.

\section{Conclusions and Outlook}

We showed how one can quantify the collectivity of atomic
excitations created by spontaneous Raman emission in
inhomogeneous ensembles, based essentially on simple
calculations for a two-level system in combination with
suitable averages over the inhomogeneous atomic spectral
distribution. We found that quite moderate detuning (of the
order of twice the inhomogeneous broadening) is already
enough to be in the regime of high collectivity, where
collective interference effects are as strong as in
homogeneous systems. This is encouraging for future
experiments in solid-state systems. It is worth mentioning
that these results are in good correspondence with the
conditions used in practice in hot gas experiments
\cite{VanderWal,EisamanPRL,EisamanNature}. Of course the
precise shape (not just the width) of the atomic
distribution has to be taken into account for any given
experiment. Note that we have assumed that the
inhomogeneous distribution is static, which is an excellent
approximation for solid-state atomic ensembles. In hot
gases collisions cause frequency changes, which become
particularly important for longer write pulses.

In this paper we have focused on the creation of an atomic
excitation in $s$, i.e. the write process of the DLCZ
protocol. The readout is a priori more complicated because
the Anti-Stokes emission exhibits collective interference
(and thus seems to be less amenable to a single-atom based
treatment), and because the Anti-Stokes photon can be
reabsorbed by the ensemble, in contrast to the Stokes
photon which couples to an essentially unpopulated
transition. However it should be possible to extend the
present approach to a detailed study of the readout as
well. The simplest case is a short, intense $\pi$ read
pulse that excites all atoms from $s$ to $e$
simultaneously, as described in the introduction. Once the
excitation has thus been transferred to the excited state
$e$, the problem is equivalent to the two-level situation
studied e.g. in Refs. \cite{Afzelius,Sangouard} and should
thus be solvable using the same techniques based on the
Maxwell-Bloch equations for inhomogeneous systems. Note
that efficient readout of excitations in $e$ is possible in
inhomogeneous, absorbing systems even in the absence of
control beams, if appropriate phase matching conditions are
fulfilled \cite{Afzelius,Sangouard}. These calculations are
typically done in a one-dimensional approximation, which
should well describe situations where write and read pulse,
Stokes photon and Anti-Stokes photon all propagate along
the same axis (in forward or backward direction). However,
it may also be possible to extend the three-dimensional
descriptions of Refs. \cite{DCZ,sorensen} to the
inhomogeneous case.

In practice, short $\pi$ pulses can be hard to implement
due to laser power limitations and due to the risk of
inducing unwanted transitions to nearby levels. It is
therefore of great interest to investigate alternative
readout schemes using chirped pulses. This requires a
detailed study of the impact of such excitation schemes on
the phases in the collective excitation. EIT effects during
the read pulse may also play a role in such a scenario
\cite{kolchinpra}. EIT in inhomogeneous systems has been
studied e.g. in Refs. \cite{kuznetsova,gorshkov}.

Let us finally note that inhomogeneous broadening can also
have desirable effects. In the context of quantum memory
protocols it should allow the efficient implementation of
temporal multiplexing \cite{Afzelius}. Such ``multi-mode
memories'' promise great speedups in the context of quantum
repeater protocols \cite{SimonMMM}. It is a fascinating
question whether a similar enhancement is possible for the
DLCZ protocol.

We thank P. Sekatski for useful comments. This work was
supported by the EU through the Integrated Project {\it
Qubit Applications} and the Swiss National Foundation
through the NCCR {\it Quantum Photonics}. C.O. was
furthermore supported by the Spanish Ministerio de
Educaci\'{o}n y Ciencia (Juan de La Cierva).


\begin{thebibliography}{99}
\bibitem{homogeneous} In homogeneous systems the
collectivity is not affected by excitation close to
resonance. However, the excitation has to be sufficiently
weak in order to avoid the emission of multiple Stokes
photons into the same spatial mode.
\end{thebibliography}
\end{document}